# Rotational Instabilities in Microchannel Flows

Saunak Sengupta[1], Sukhendu Ghosh[2], Sandeep Saha[3] and Suman Chakraborty[1]

[1]Department of Mechanical Engineering, Indian Institute of Technology, Kharagpur 721302, India

[2]NLPC Unit, Faculty of Sciences, Université Libre de Bruxelles, Brussels 1050, Belgium

[3]Department of Aerospace Engineering, Indian Institute of Technology, Kharagpur 721302, India

## Abstract

Mixing in numerous medical and chemical applications, involving overly long microchannels, can be enhanced by inducing flow instabilities. The channel length, is thus shortened in the inertial microfluidics regime due to the enhanced mixing, thereby rendering the device compact and portable. Motivated by the emerging applications of lab-on-a-CD based compact microfluidic devices, we analyze the linear stability of rotationally actuated microchannel flows commonly deployed for biochemical and biomedical applications. The solution of the coupled system of Orr-Sommerfeld (OS) and Squire (SQ) equations yields the growth rate and the neutral curve of the two types of instabilities: (i) the *Tollmien-Schlichting* (TS) wave and (ii) the *Coriolis force-driven instability*. We report the existence of four distinct unstable modes (Modes $I-IV$) at low Reynolds numbers $\left(O(\text{Re}) \sim 100\right)$ of which *only* the existence of Mode $I$ is previously known for the present flow configuration. Furthermore, Modes $I$ and $II$ exhibit competing characteristics, signifying that Mode $II$ can also play *an important role* in the transition to turbulence. Modes $III$ and IV have relatively lower growth rates, but the associated normal velocity has an oscillatory nature near center of the channel. Thus, we infer that Modes $III$ and $IV$ might cause strong mixing locally by virtue of strong velocity perturbation in proximity to the interface; a scenario plausible if the channel is too short to allow for the amplification of Modes $I$ and $II$. We quantify the potential of all the modes to induce such localized mixing near the interface using the notion of penetration depth. We also present an instability regime map obtained from the parametric study over a range of Reynolds numbers, Rotation numbers, streamwise and spanwise wave numbers, to assist the design of efficient microchannels. Further insight into the mechanism of energy transfer, drawn from the evaluation of kinetic-energy budget, reveals that the Reynolds stress first transfers energy from the mean flow to the streamwise velocity fluctuations. The Coriolis force, thereafter, redistributes the axial momentum into spanwise and wall-normal directions, generating the frequently observed roll-cell structures. A qualitative comparison of our predictions with reported experiments on roll-cells indicates co-existence of Modes $I$ and $II$.

**1. Introduction**

Mixing processes at the microscale play a key role in the progress of several chemical and biological reactions such as DNA hybridization (Wei et al. 2005), bioreactors (Li 2008) and immuno-assays (Hatch et al. 2001). Such systems are typically characterized by low Reynolds numbers $\text{Re} \sim \text{O}(1), \text{Re} := \text{U}_m D_h \nu^{-1}$, where the mean axial velocity is $\text{U}_m$, the hydraulic diameter of the microchannel is $D_h$ and $\nu$ is the viscosity of the fluid. Mixing is typically diffusion dominated (*i.e.* Peclet number, $Pe \sim \text{O}(100)$) in a microchannel, and therefore a slow process. The mixing length varies linearly with the Peclet number ($Pe$) (Stroock 2002), and therefore the resulting design for micro channels is often beyond practically acceptable requirements (Liu et al. 2000),(Steigert et al. 2005). The hastening of the mixing process is thus important for further development of microchannel devices.

Enhanced mixing at such small scales is achieved by a plethora of techniques (Erdogan & Chatwin 1967; Berger et al. 1983; Jiang et al. 2004; Sudarsan & Ugaz 2006), mostly based on introducing a secondary flow caused by inertial forces. Rotationally actuated micro-fluidic devices, like Lab-on-a-CD platform (see figure 1), are inherently adept at generating secondary flows by virtue of the Coriolis force. The advantages are many (Burger et al. 2012), for instance: (i) versatility with specimen irrespective of the fluid properties like viscosity, conductivity, (ii) uncomplicated geometry unlike Dean force-based devices which require specialized micro-fabrication (Bertsch et al. 2001; Kim et al. 2004), (iii) simple rotor actuation mechanism, thus making the device ideal for low cost environs. Configurations, where the Coriolis force overwhelms the centrifugal force, are used in practice to enhance micro-mixing by triggering instabilities (Madou et al. 2006; Madou et al. 2001; Ducrée et al. 2007; Chakraborty & Chakraborty 2010). Hence, a clearer understanding of the destabilizing effect of the Coriolis force in a CD-based platform can lead to more precise and effective designs. In this context, our study lies at the intersection of inertial microfluidics and linear stability theory.

Figure 1 illustrates a rotating microchannel flow where the incepted instability leads to intense mixing. Experiments and subsequent theoretical calculations (Chakraborty et al. 2011) reveal that the ratio of Coriolis force to the centrifugal force $(\beta)$ governs the extent of mixing achieved in such flows. The authors report that molecular diffusion dominates at low rotational speed because Coriolis force is lower than the centrifugal force. Mixing due to a streaky instability gets prominence when the rotational speed is such that $\beta$ exceeds 1.35, and the transverse Coriolis force is large enough to setup a secondary flow. Numerical simulations of Roy et al. (2013) indicate that increasing the channel aspect ratio leads to a non-monotonic behavior in the critical rotational Reynolds number $Re_{\omega,cr}$ at which the secondary flow sets in, with the lowest corresponding to a square channel. These observations raise question as to how the Coriolis force destabilizes the flow.

In order to understand and answer the question better, we briefly review the linear stability of rotating microchannel flows. Plane Poiseuille flow, known to be stable until Reynolds number $3848.15$ (Orszag 1971), when subjected to even a minor spanwise perturbation, becomes unstable at very low Reynolds number ($\text{Re}_{cr} \approx 88.53$ (Lezius & Johnston 1976) and $100$ (Alfredsson & Persson 1989)) in the presence of spanwise rotation. The transitional and turbulent rotating channel flows are marked by the appearance of streaky or streamwise oriented roll-cell structures (Chakraborty et al. 2011; Lezius & Johnston 1976; Matsson & Alfredsson 1990; Grundestam et al. 2008; Kristoffersen & Andersson 1993) which have been attributed to the Coriolis force. Indeed, the wavelength of the *steady* roll-cell instability can be predicted using linear stability theory (Matsson & Alfredsson 1990). Intriguingly, the roll-cell structures are gradually eliminated with an increase in the rotational speed (Alfredsson & Persson 1989; Wall & Nagata 2006).

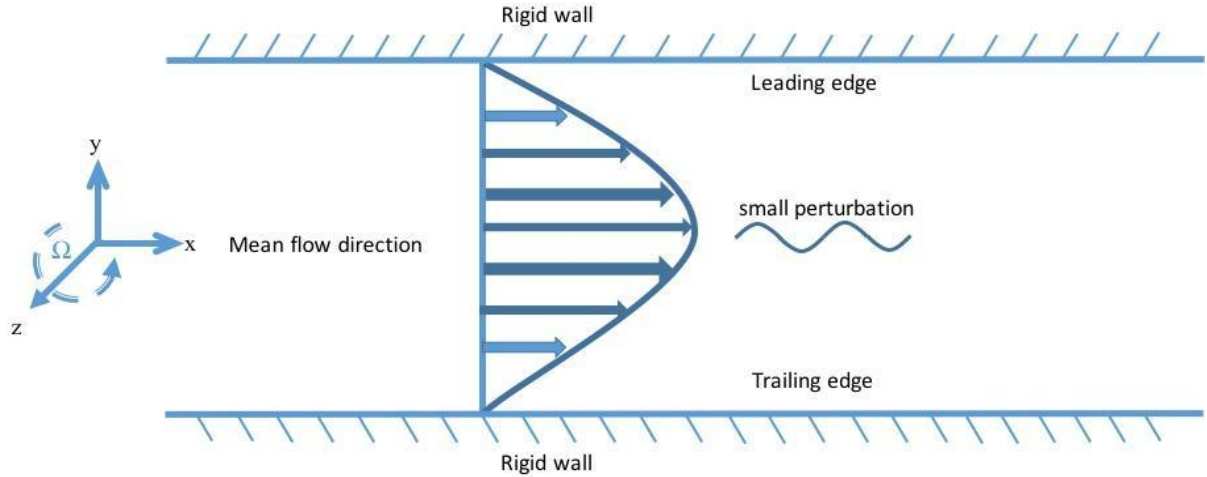

Figure 1. Schematic of the Lab-on-CD device showing transition to turbulence due to roll-cells.

Despite the experimental and theoretical investigations by researchers, a complete understanding of the Coriolis force induced roll cell instability (*i.e.* Coriolis instability) has failed to emerge and there are questions yet to be answered. For instance, how does a variation in the rotation rate affect the growth of roll-cells, or how does fluid viscosity influence the neutral curve for the Coriolis instability, can there be multiple unstable modes and can they co-exist? Is there a competition between the multiple unstable modes emerging from the Coriolis instability? Furthermore, the precise energy transfer mechanism, which excites the roll-cells, is also unclear. The development of design guidelines, hitherto lacking, for microchannels using Lab-on-CD technique, rests on our understanding of such instabilities. In order to answer these fundamental queries and develop a theoretical understanding of the effect of Coriolis force, we conduct a linear stability analysis of a rotationally actuated flow.

Our analysis reveals the existence of four unstable modes $(Type\, I - IV)$, triggered by the Coriolis force, at various Reynolds and Rotational numbers. The $Type - I$ and $II$ modes are regularly spaced roll-cells exhibiting mode competition. $Type - III$ and $IV$ modes have a twisted roll-cell like structure and have lower growth rates although they are more effectively participating in micromixing process. Examination of the kinetic energy distribution provides insight into the physical mechanism by which the roll-cells extract energy from the mean flow. We make a qualitative comparison with the experimental results of Alfredsson & Persson (1989), with the predictions of the theoretical model. The comparison reveals that roll-cell like structures can be predicted by the model, but more importantly indicates that there may be multiple unstable modes co-existing in the flow. The presence of multiple modes can alter the transition path and thereby have significant impact on the mixing process. A regime diagram delineates about the modes induced by the Coriolis force at small rotation number to guide the development of precisely designed microchannels.

The article is organized as follows: In section 2, we present the mathematical formulation of the problem and the linearized Navier-Stokes equations are discussed. In section 3 we present the stability results. Then, in section 4, we present the energy budget calculation and energy transfer mechanism for the system under consideration. In section 5, we compare our numerical results with experimental results of Alfredsson & Persson (1989). We also provide design guidelines for microchannels in section 6. We finally summarize our calculations in section 7.

## 2. Mathematical Formulation

### 2.1 Base flow

The linear stability of a Newtonian, incompressible, pressure-driven laminar channel flow with a spanwise system of rotation (Alfredsson & Persson 1989; Lezius & Johnston 1976; Wallin et al. 2013) is considered. The fluid has density $\rho$ and kinematic viscosity $\nu = \mu/\rho$. A schematic diagram of the flow in a Cartesian coordinates is illustrated in figure 1. The co-ordinate axes $x$ and $y$ are placed along the streamwise and cross-streamwise flow directions, respectively, and the entire system is rotating about the $z$ axis. The dimensional form of the governing equations is:

$$\nabla \cdot \vec{u} = 0 \tag{1}$$

$$\frac{\partial \vec{u}}{\partial t} + (\vec{u}.\nabla)\vec{u} = -\frac{1}{\rho}\nabla p_m + \nu\nabla^2\vec{u} - \frac{2}{\rho}(\Omega \times \vec{u}) \tag{2}$$

where $\vec{u}$, $p$, $p_m$ $\left(p_m = p - \rho\,\Omega^2(x^2 + y^2)\right)$ $\rho$, $\nu$ and $\Omega$ are the velocity vector, static pressure modified pressure, density, kinematic viscosity of the flow system and angular speed of the rotating channel, respectively. The last term in the momentum equation signifies the effect of the Coriolis force. We use the width of the channel $(D_h)$ as the length scale, the average velocity as the velocity scale.

The problem admits a steady, parallel, fully developed base-flow described by the non-dimensional expression $U(y) = 3/2(1-y^2)$. A further discussion on the modified pressure distribution in such systems is provided by Liu et al. (2008).

## 2.2 The linear stability equations

Following the modal linear stability approach, we perturb the base state velocity $(U(y),0,0)$ and modified base pressure $p_m$ with an infinitesimal three-dimensional disturbance $(u,v,w,p)$, and linearize the governing equations with reference to the disturbance. The resulting perturbed flow is governed by the following set of linearized mass and momentum balance equations in the reference frame of the microchannel,

$$\frac{\partial u}{\partial x} + \frac{\partial v}{\partial y} + \frac{\partial w}{\partial z} = 0 \tag{3}$$

$$\frac{\partial u}{\partial t} + U\frac{\partial u}{\partial x} + v\frac{dU}{dy} = -\frac{1}{\rho}\frac{\partial p}{\partial x} + \nu\nabla^2 u + 2v\Omega \tag{4}$$

$$\frac{\partial v}{\partial t} + U\frac{\partial v}{\partial x} = -\frac{1}{\rho}\frac{\partial p}{\partial y} + \nu\nabla^2 v - 2u\Omega \tag{5}$$

$$\frac{\partial w}{\partial t} + U\frac{\partial w}{\partial x} = -\frac{1}{\rho}\frac{\partial p}{\partial z} + \nu\nabla^2 w \tag{6}$$

The no-slip and no-penetration boundary conditions are applied at the walls ($y = \pm D_h$),

$$u(\pm D_h) = v(\pm D_h) = w(\pm D_h) = 0 \tag{7}$$

Further, the equations (3)-(6) are recast into a set of coupled differential equations in terms of wall-normal velocity $v$ and wall-normal vorticity $\eta := \partial u/\partial z - \partial w/\partial x$. In order to analyze the linear stability of the system, we invoke the normal-mode assumption,

$$\begin{pmatrix} v(x,y,z,t) \\ \eta(x,y,z,t) \end{pmatrix} = \begin{pmatrix} \tilde{v}(y) \\ \tilde{\eta}(y) \end{pmatrix} \exp\left[i(k_x x + k_z z - \omega t)\right] \tag{8}$$

Where ~ denotes eigen function of wall normal velocity and vorticity and $k_x, k_z$ are the streamwise and spanwise wave numbers, and $\omega$ is the frequency of the disturbances. Finally, we have,

$$\frac{\partial}{\partial t}\begin{pmatrix} \Delta \tilde{v} \\ \tilde{\eta} \end{pmatrix} = \begin{pmatrix} L_{OS} & F_C \\ L_C & L_{SQ} \end{pmatrix}\begin{pmatrix} \tilde{v} \\ \tilde{\eta} \end{pmatrix} \tag{9}$$

Wherein, the operators,

$$L_{OS} = -ik_x U \Delta + ik_x U'' + \frac{\Delta^2}{Re}; F_C = -iRo k_z \tag{10}$$

$$L_C = -ik_z(U' - Ro); L_{SQ} = -ik_x U + \frac{\Delta}{Re} \tag{11}$$

$$\tilde{v}(\pm 1) = D\tilde{v}(\pm 1) = \tilde{\eta}(\pm 1) = 0. \tag{12}$$

with $D \triangleq d/dy$, $\Delta \triangleq D^2 - k_x^2 - k_z^2$, Reynolds number $Re = UD_h/\nu$ and the rotation number $Ro = \Omega D_h / U$, respectively. The matrix equation (9) solved together with the boundary conditions (12) is an fourth order eigen value problem solved using a Chebyshev spectral collocation method following the numerical procedure outlined by Trefethen (2000).

## 3. Results and Discussions

### 3.1 Validation of the mathematical model

We have validated our numerical code against the results of Wallin et al. (2013) and (Alfredsson & Persson 1989). Wallin et al. (2013) studied the laminarization of a turbulent rotating plane channel flow and reported the stability of crossflow modes at large Reynolds numbers. Figure 2(a) shows that our code is able to reproduce the neutral stability curve of Wallin et al. (2013) for the crossflow modes in $k_z - Ro$ plane. Similarly, the neutral stability curve in $k_z - Re$ plane predicted by Alfredsson & Persson (1989) using spatial stability theory for the stationary roll-cells (for zero frequency) is also captured by our code in by setting $\omega_i = 0$ figure 2(b).

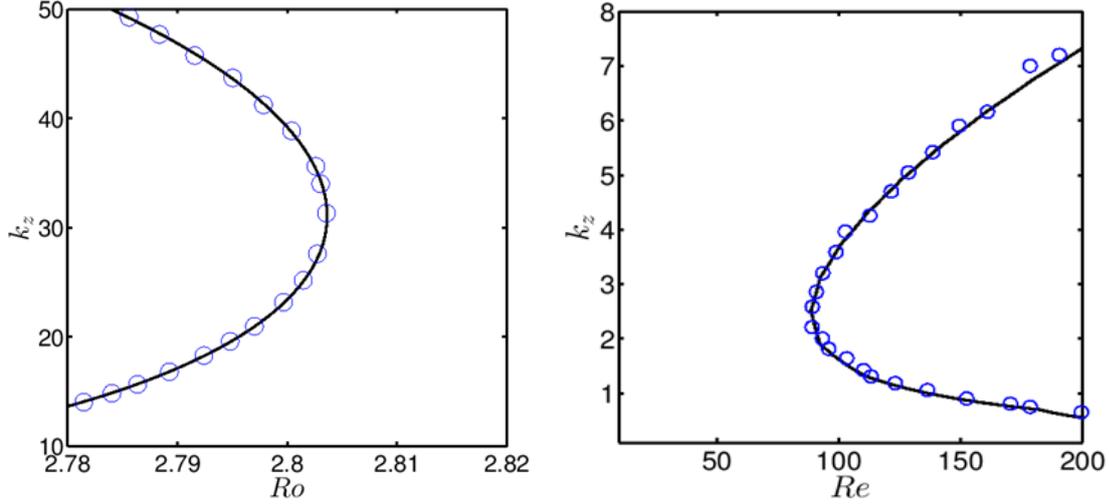

Figure 2. .Neutral stability curves for rotating channel flow: (a)comparison with Wallin et.al. (2013), (b) comparison with Alfredsson & Persson (1989). In figure (a) $Re = 10800, k_x = 0.0$; symbols are used to show the results of Wallin et al., (2013), and in (b) $Ro = 0.5, k_x = 0.0$ with Alfredsson & Persson (1989).

**3.2 Modal instabilities and regime diagram**

In congruence with the earlier results (Lezius & Johnston 1976, Alfredsson & Persson 1989), we find the existence of one unstable mode for low Reynolds numbers $\left(\sim O(100)\right)$, which we refer to as $Type-I$ instability. However as the Reynolds number is increased to reach approximately 200, we find the emergence of another unstable mode that we refer to as $Type-II$ instability. The $Type-II$ instability has not yet been reported for this configuration at low Reynolds numbers.

In addition, we have found two more unstable modes, namely $Type-III$ and $Type-IV$, which too have no precedents in literature. Figure 3(a) presents the eigenvalue spectrum for Reynolds number $Re = 1228.0$ and rotation number $Ro = 0.15$, wherein all the four modes $(Type\ I-IV)$ are unstable and may co-exist in the flow. $Type-I$ mode has the highest growth rate $(\omega_i)$ followed by $Type-II$, $Type-III$ and $Type-IV$ modes with nearly equal phase speeds. In figure 3(b) we show the eigenspectrum for $k_x = 0.0$, where $(Type\ I-IV)$ modes exist but the phase speed is zero, confirming that the roll cells formed are essentially standing waves (Alfredsson & Persson 1989). Furthermore, all the four unstable modes $(Type\ I-IV)$ are driven by the *Coriolis* force and are not shear instabilities. This conclusion is drawn from figure 3(c) where we observe that the unstable modes cease to exist when Coriolis force is removed by setting $Ro = 0$.

The eigenfunction of normal velocity and vorticity associated with $Type\ I-IV$ modes is shown in figure 3(d).

We observe that the eigenfunctions $[v,\eta]$ for the four unstable modes have a very distinct character in the wall-normal direction in figure 3(d). The length-scale in the wall-normal direction decreases, alternatively the wall-normal wavenumber increases, as we move from modes $Type-I$ to $Type-IV$. We expect that the associated viscous dissipation term might increase too, which would usually lead a reduction in the growth rate as observed in figure 3(b). We show that indeed the viscous dissipation is enhanced progressively from $Type-I$ to $Type-IV$ modes in table 1 of section 4 where we discuss the magnitude of the various terms of the kinetic energy equation.

Another interesting aspect of these instabilities is the region over which the instability is present. We observe that the eigenfunction is largely restricted to the upper half of the channel for all the modes while the tendency to penetrate into the lower half increases from $Type-I$ to $Type-IV$ with the wall-normal wavenumber. To understand the effect of Coriolis force we plotted the eigenfunctions in figure 3(e) for clockwise rotation keeping rest of the parameters identical as in figure 3(d). We observe that the eigenfunctions are restricted to the lower half of the wall. This confirms that the disturbances largely depend on the direction of rotation, i.e. Coriolis force.

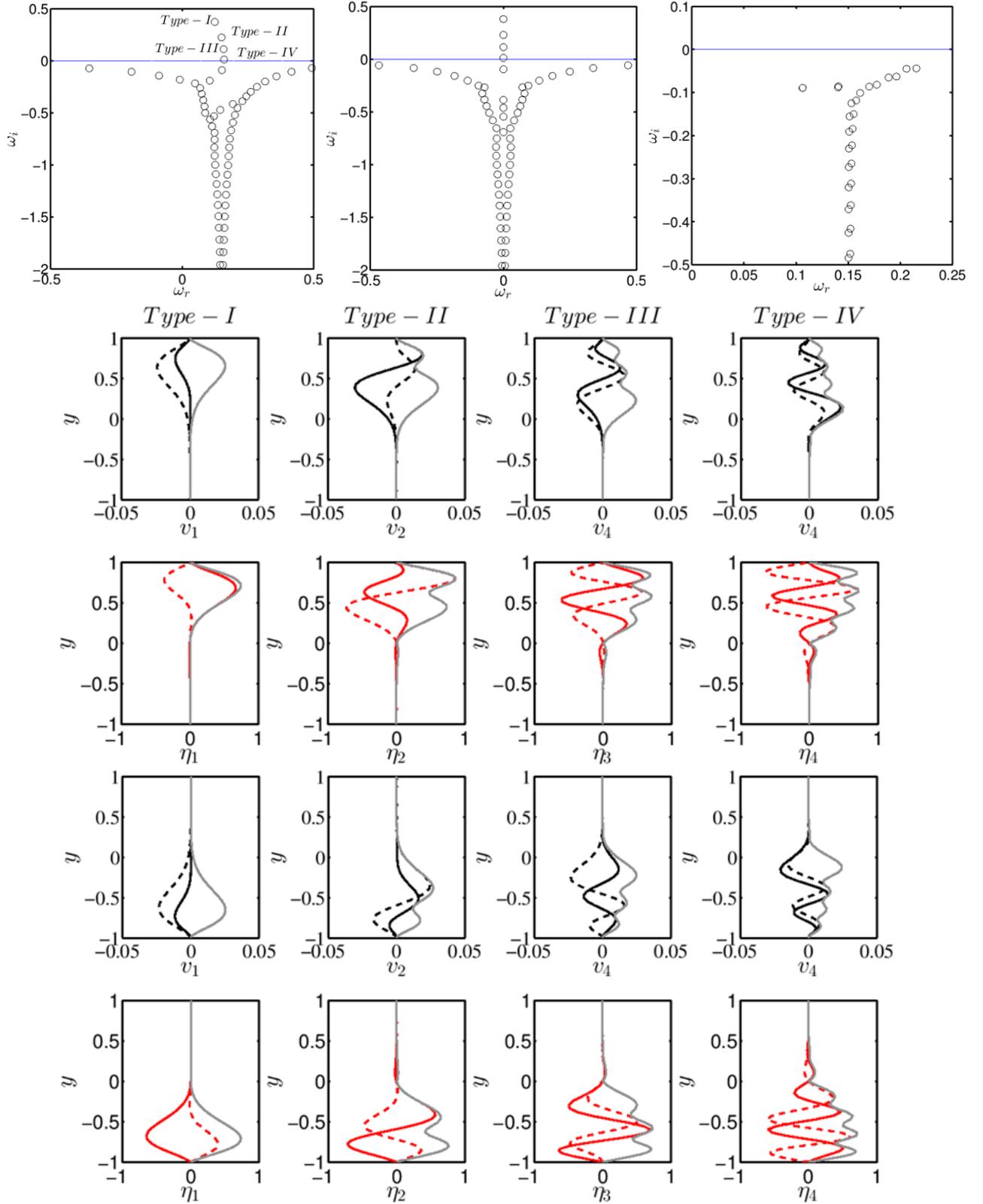

Figure 3. (a) Eigenvalue spectrum at $Re = 1228, Ro = 0.15$ and $k_x = 0.15, k_z = 6.5$, (In the inset four unstable eigenmodes). (b) Eigenfunctions for the unstable eigenmodes $Type\ I-IV$, corresponds the wall normal velocity $(\tilde{v})$. Solid, dashed and gray lines present real, imaginary and absolute values of eigenfunctions respectively. (c) Eigenvalue spectrum at $Re = 1228, Ro = 0.0$ and $k_x = 0.15$, $k_z = 6.5$ (All modes are stable). Eigen functions for above mentioned parameters for (d) anticlockwise rotation & (e) clockwise rotation.

### 3.2.1 Structure of Type I-IV modes

We now examine the effect of streamwise and spanwise wavenumber variation on the perturbation velocity field. Figures 4(a) to (d) show the structure of the perturbation field for $Type-I$ to $Type-IV$ modes respectively. We observe that roll-cells corresponding to the instabilities are essentially streamwise oriented vortices and exhibit greater number of twists in the wall-normal direction as we move from $Type-I$ to $Type-IV$ modes. However, the twists vanish when we set $k_x = 0$, as can be seen on the right pane of figures 4(a) to (d) where the disturbance form a two-dimensional standing wave. The twisting of the roll-cells is not entirely obvious from the flow visualization images reported by Alfredsson and Persson because they have obtained the images in a particular plane. However the three-dimensional disturbance flow field exhibits clearly the twisting nature of the roll-cells. The greater number of twists indicates that $Type-III$ and $Type-IV$ modes might be more effective in causing mixing. We also notice that the twisting modes penetrate less into the lower half of the channel compared to the two-dimensional modes. Similar observations about the presence of two-dimensional roll-cells near the channel centreline can be made in the figures of the secondary flow reported by (Speziale & Thangam 1983).

### 3.3 Reynolds and rotation number dependence

We analyse the effect of varying Reynolds number on the neutral stability curve for the fastest growing instability in the $k_z - Ro$ plane as shown in figure 5(a) for the two-dimensional disturbances by setting $k_x = 0$. The $Type-I$ mode is the dominant instability for two-dimensional cases and we observe that there exists an island like unstable region in the $k_z - Ro$ plane. The size of this region gradually increases as the Reynolds number increases. Furthermore, the inset of figure 5(a) reveals that the critical $Ro \approx O(0.01)$ is very low and progressively reduces as the Reynolds number increases while the critical spanwise wavenumber remains almost constant. The observations indicate that the $Type-I$ mode is essentially an inviscid instability mechanism, unlike the Tollmien-Schlichting mode. Indeed, the gradient Richardson number for rotating flows, $B^* = Ro(Ro - U') > 0$ ((Bradshaw 1969), (Wallin et al. 2013)) determines the stability for inviscid flows. Our analysis is in agreement with prediction of the invscid criterion because a reduction in magnitude of the viscous terms yields a neutral curve with the limiting values of the critical $Ro$ approaching the 0 and $U'$. We refer to the lower bound as the sub-critical $Ro$ and the upper bound as the super-critcal $Ro$ and for most microfluidics applications the former is of concern. We report the variation of sub-critical $Ro$ with Reynolds number in section 6 dedicated to micro-channel design guidelines.

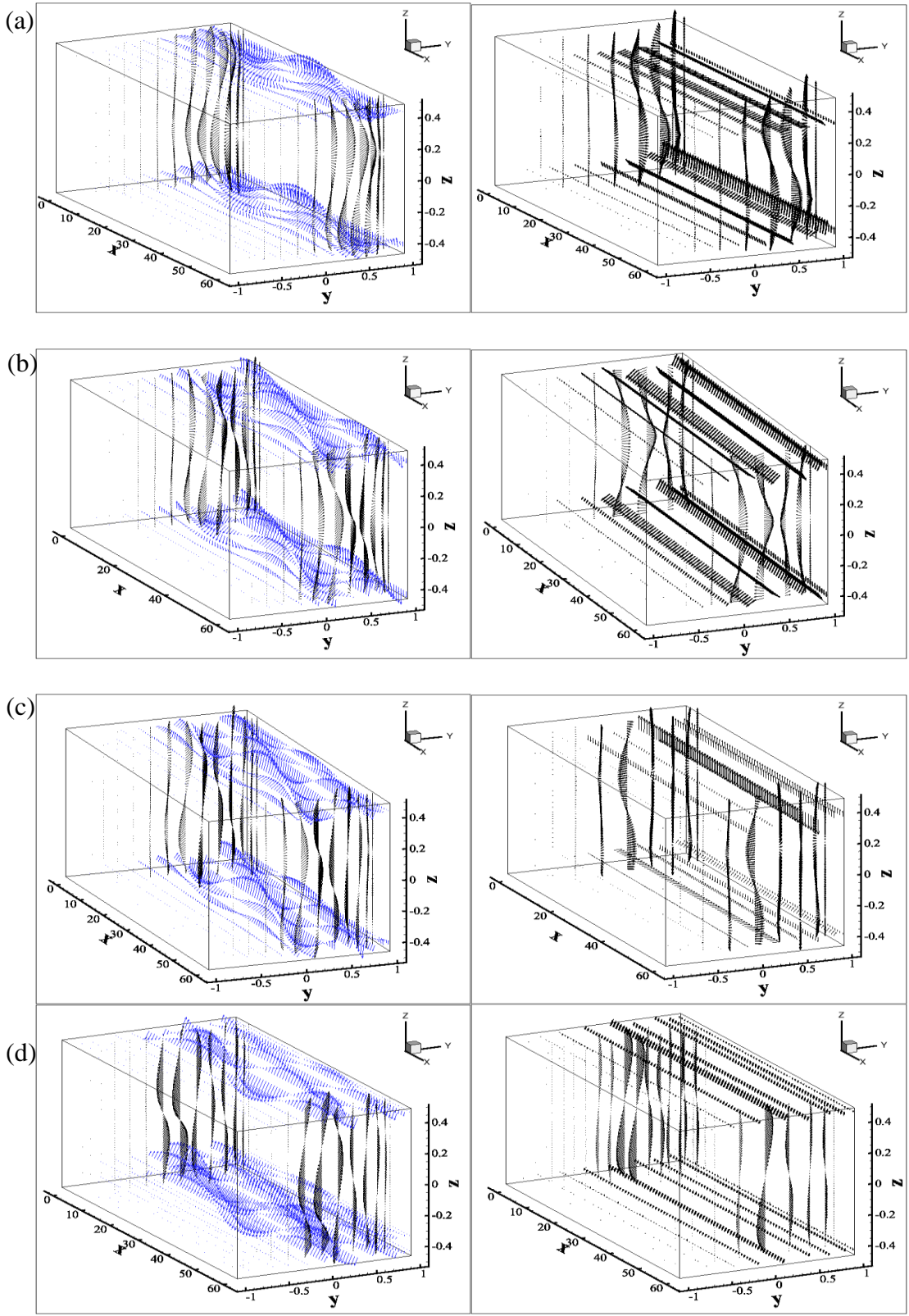

Figure 4. The structure of roll cells formed due to the existence of (a) $Type-I$, (b) $Type-II$, (c) $Type-III$ & (d) $Type-IV$ instability. The used parameters are $k_x = 0.15$, (left column), $0.0$, (right column) $k_z = 7.5, Re = 1228.0,$ and $Ro = 0.15$.

However, as we increase the streamwise wavenumber from $k_x = 0$ to $k_x = 1.2$, we observe that the neutral stability curve has a develops a discontinuity in the gradient as shown by the symbol $A$ region of figure 5(b). The discontinuity vanishes if we increase the streamwise wavenumber such that $k_x \geq 1.9$. Faller (1991)(Faller 1991) also reported similar type of discontinuity for the flow over a rotating disk. The discontinuity arises because of mode competition between $Type-I$ and $Type-II$ modes. In essence, the growth rate of the $Type-II$ mode dominates over the growth rate of $Type-I$ mode for a range of streamwise wave numbers. Therefore, a discontinuity in the gradient arises because of the intersection of the neutral stability curve of $Type-I$ and $Type-II$. This is a novel finding and could potentially be beneficial in triggering early transition in microfluidic channels by introducing three-dimensional perturbations.

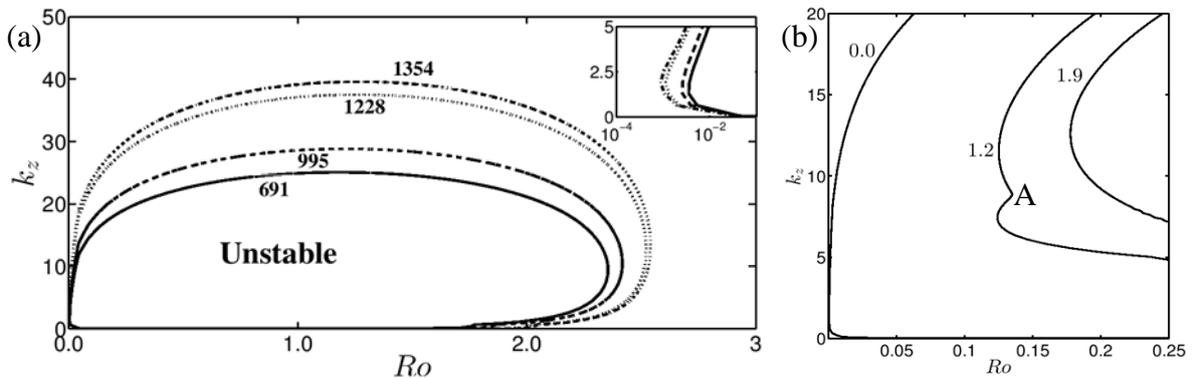

Figure 5. (a) Neutral curves for different Reynolds number $Re$ (with $k_x = 0$). An island of unstable zone has been observed (Onset of the instability is shown by the inset figure). (b) Neutral curves as a function of streamwise wave number ($k_x = 0, 1.2, 1.9$) at $Re = 1228$.

### 3.3.1 Mode competition

The growth rate of $Type\ I-IV$ instabilities over a range of spanwise wave number $k_z$ for $Re = 1228.0$ and $Ro = 0.15$ for two different values of streamwise wave number $(k_x = 0.15, k_x = 1.2)$ is shown in figures 6 (a) and (b) respectively. For three-dimensional disturbances, the maximum growth rate of all the four modes, shown by circles, lies in a narrow band of $k_z$ $(7.9242 \leq k_z \leq 9.4091)$. The growth rate of $Type-I$ instability dominates over all other instabilities for three-dimensional disturbance. The same is observed if we change the above said parameters. At higher streamwise wavenumber, the three dimensional instabilities are in contrast to lower streamwise disturbances $(k_x)$; in figure 9(b) we observe mode competition between $Type-I$ and $Type-II$ for $k_x = 1.2$. Notably, $Type-II$ mode dominates over $Type-I$ mode for a set of $k_z$ value, and thereby can potentially contribute to the roll-cell instability reported in the experiments previously (Alfredsson & Persson). Figure 9(b) also confirms the fact that by increasing streamwise wave number we can suppress the $Type-III, Type-IV$ instability. A higher streamwise wave number also reduces the growth rate of $Type-I$ and $Type-II$ modes, and perhaps could be beneficial in controlling the extent of mixing in centrifugal microfluidic devices.

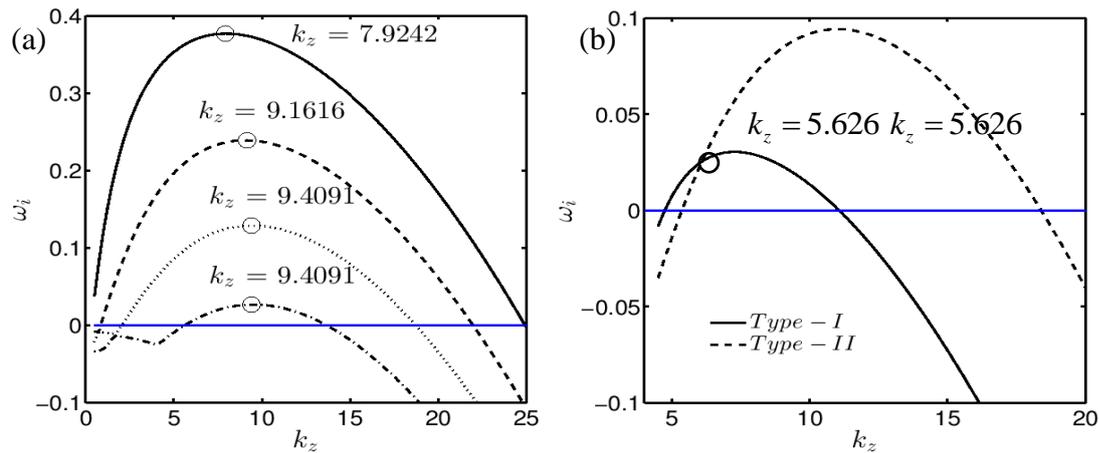

Figure 6. Growth rate curves for $k_x$ (a) $0.15$ and (b) $1.2$. Other parameters are $Re = 1228.0, Ro = 0.15$ respectively. In figure (a) $Type-I$ solid line represents, dashed line $Type-II$, dotted line $Type-III$ and dash dot line $Type-IV$.

## 4. Energy Budget and Instability Mechanism

In the previous subsection we found the existence of mode competition wherein $Type-II$ instability can dominate over $Type-I$. The observation is however counterintuitive because the eigenfunction of $Type-II$ instability has a higher wavenumber in the wall-normal direction compared to $Type-I$ suggesting that viscous dissipation might be a dominant phenomenon for $Type-II$ instability. Hence in order to understand how does $Type-II$ instability overwhelm the growth of $Type-I$ instability we examine the kinetic energy equation following the procedure described by (Faller 1991). In order to proceed we must first identify the mechanism by which the energy of the base flow is passed over to the roll-cells and the role of Coriolis force in the process of energy transfer. The kinetic energy equation can be expressed as in equations (13) and (14),

$$\underbrace{\frac{\partial}{\partial t}\iiint \frac{1}{2}u'^2 dxdydz}_{I} = \underbrace{-\iiint u'v'\frac{dU}{dy}dxdydz}_{II} + \underbrace{\frac{1}{Re}\iiint \left[\left(\frac{\partial u'}{\partial x}\right)^2 + \left(\frac{\partial u'}{\partial y}\right)^2 + \left(\frac{\partial u'}{\partial z}\right)^2\right]dxdydz}_{III} \quad (13)$$
$$+ \underbrace{Ro\iiint u'v'dxdydz}_{IV},$$

$$\underbrace{\frac{\partial}{\partial t}\iiint \frac{1}{2}(v'^2 + w'^2)dxdydz}_{V} = \underbrace{\frac{1}{Re}\iiint \left[\begin{array}{c}\left(\left(\frac{\partial v'}{\partial x}\right)^2 + \left(\frac{\partial v'}{\partial y}\right)^2 + \left(\frac{\partial v'}{\partial z}\right)^2\right) \\ + \left(\left(\frac{\partial w'}{\partial x}\right)^2 + \left(\frac{\partial w'}{\partial y}\right)^2 + \left(\frac{\partial w'}{\partial z}\right)^2\right)\end{array}\right]dxdydz}_{VI} \quad (14)$$
$$- \underbrace{Ro\iiint u'v'dxdydz}_{IV}.$$

wherein the kinetic energy $(K.E.)$ of the disturbance field is decomposed into (a) kinetic energy of the streamwise component $(K.E_u)$ and (b) kinetic energy of the roll-cells $(K.E_{vw})$ in order to highlight the role of Coriolis force. In eq. (13), the term $I$ represents the rate of change of $K.E.$ originating from the $x$-component of the disturbed flow and the term $V$ represents the rate of change of $K.E.$ induced by overturning cells. Following Faller's (1991)(Faller 1991) notation, we denote term $I$ as $K.E_u$ and the term $V$ as $K.E_{vw}$ and therefore, the total amount of disturbance kinetic energy inside the flow is $K.E. = K.E_u + K.E_{vw}$. In equation (13), term $II$ represents the transfer of energy from the base flow to the perturbed flow, known as the Production/Reynolds stress term. Term $IV$ appears in eq. (13) and (14) with a change in sign and illustrates the energy transfer because of the Coriolis force from production term to the kinetic energy of the roll-cells. Terms $III$, $VI$ in equations (13) and (14) denote the viscous dissipation terms which are always negative. The flowchart in figure 7 summarizes the flow of energy from base flow to the overturning cells for all the four unstable modes. The Production term extracts energy from the mean flow and transfers partially to $K.E_u$ and the Coriolis term and the rest is lost to viscous dissipation. The Coriolis term thereafter transfers, all the energy from production term, to $K.E_{vw}$ and viscous dissipation. The role of the Coriolis term is thus clarified by the decomposition of the total kinetic energy equation into equations (13) and (14).

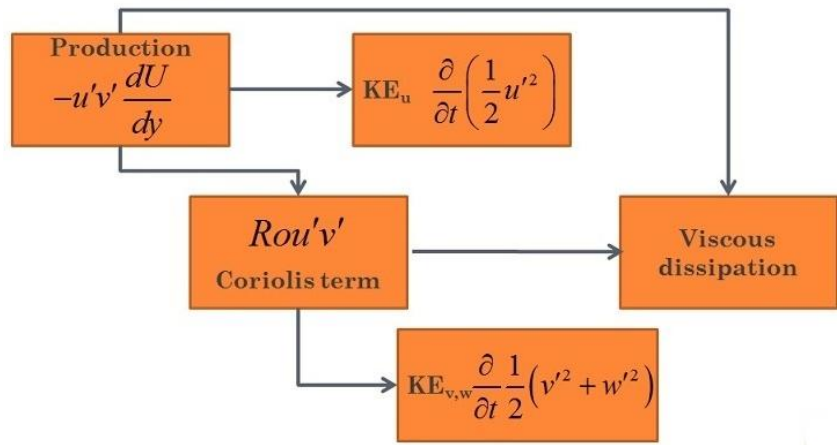

Figure 7. Flow chart of energy transfer for the channel flow in a rotating platform.

In table 1 and fig. 8 (a)-(d) we have shown the terms of the kinetic energy equation and the corresponding variation in the wall normal direction for $Types\ I-IV$ instabilities. The $Type-I$ instability extracts energy from the base flow most efficiently using the production term compared to the other modes (see fig. 8(a)-(d)) and thus is able to transfer a significant fraction to the Coriolis term which in turn sustains the overturning cells and therefore has the highest growth rate. This mechanism of energy transfer is observed in the other instabilities too. However, as move from $Type-II$ to $Type-IV$ instability the production term diminishes and the dissipation terms $Diss_u$, $Diss_{vw}$ increase in magnitude due to the higher wall-normal wavenumber (see fig. 8(b)-(d)), progressively suppressing the growth rate. It is also interesting to note that the higher modes like $Type-III$ and $IV$ persist deeper into the channel compared to $Type-I$. This behaviour is reminiscent of the notion of penetration depth introduced by Jacobs & Durbin (1998) and subsequently explained by ZAKI & SAHA (2009).

Further, it appears that the leading order terms of magnitude, in table. 1 are the production and $K.E_u$ for $Type-I$ instability while $Diss_u$ is a higher order term; however, for $Type-IV$ instability the production and $Diss_u$ contribute to the leading order in the energy equation. This implies that relative to $Type-I$, the $Type-IV$ instability has a Coriolis force which is able to transfer a greater portion of the production term to the overturning cells and the kinetic energy in the streamwise velocity reduces (see fig. 8(a)–(d)). Indeed, all the unstable modes amplify only because of the Coriolis term. In absence of the Coriolis term no unstable mode exists for spanwise perturbations for the given rotation and Reynolds number as observed in Fig. 3(c). We thus conclude that all the modes are driven by the Coriolis force.

| Mode | Growth rate $(\omega_i)$ | Production | $K.E_u$ | Coriolis | $K.E_{vw}$ | $Diss_u$ | $Diss_{vw}$ |
|---|---|---|---|---|---|---|---|
| $Type-I$ | 0.37174559 | 10.750884 | 8.472732 | 1.628246 | 0.690161 | 1.448784 | 0.1391412 |
| $Type-II$ | 0.22380689 | 7.479152 | 4.798398 | 1.385525 | 0.5056563 | 1.965839 | 0.2092014 |
| $Type-III$ | 0.11093088 | 6.328904 | 2.636803 | 1.289553 | 0.3260593 | 3.022627 | 0.3435847 |
| $Type-IV$ | 0.01100349 | 5.006776 | 0.4153594 | 1.057479 | 0.05830904 | 4.040684 | 0.4924215 |

Table 1. Components of the kinetic energy equation for $Types\ I-IV$ instabilities for $Re=1228, Ro=0.15$ and $k_z=7.9, k_x=0.15$.

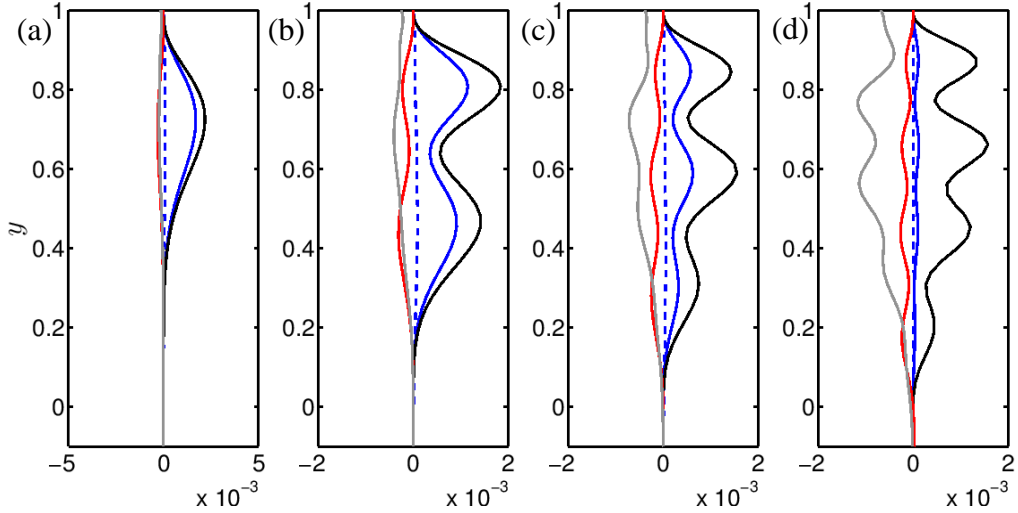

Figure 8. Energy budget for four different mode (a) $Type-I$, (b) $Type-II$, (c) $Type-III$ and (d) $Type-IV$. Black lines represents production term, blue solid and dashed lines represents $K.E_u$ and $K.E_{vw}$ respectively, red lines Coriolis terms and gray lines are for dissipation terms.

## 4.1 Kinetic energy budget for mode competition

We now examine the energy transfer mechanism for competing modes in Tables 2(a), 4(b) and 2(c) corresponding to dominance of $Type-I$ over $Type-II$, nearly equal growth rate of $Type-I$ over $Type-II$, and dominance of $Type-II$ over $Type-I$ respectively. In table (a), we note that the ratio of $K.E_{vw}$ to Coriolis term is higher for $Type-I$ compared to $Type-II$. In table (b), where mode crossing appears, the ratio of $K.E_{vw}$ to Coriolis term is almost equal for both the modes $(Type-I \& II)$, even though the amount of energy transferred by the production term to the Coriolis term and consequently to $K.E_{vw}$ may differ in magnitude. In table (c), where $Type-II$ dominates, we observe the production of $Type-I$ is lower compared to $Type-II$ but the Coriolis term is larger in $Type-I$ than $Type-II$. Consequently the $K.E_{vw}$ is also smaller for $Type-II$ compared to $Type-I$ in absolute terms and yet $Type-II$ instability dominates over $Type-I$. However, the Coriolis term contributes relatively more to $K.E_{vw}$ of $Type-II$ as compared to $Type-I$ and is therefore a better measure for predicting the growth rate. It is also worth noting that the $Type-II$ instability is distinct from that of the $Type-II$ mode reported by Faller (1991)(Faller 1991) because the direction of energy transfer of the Coriolis terms is reversed.

|     | Growth rate ($\omega_i$) | Production | $K.E_u$ | Coriolis | $K.E_{vw}$ | $Diss_u$ | $Diss_{vw}$ |
|-----|--------------------------|------------|---------|----------|------------|----------|-------------|
| (a) | 0.013778                 | 42.8054    | 3.07160 | 13.0694  | 1.10248    | 28.8522  | 9.7663      |
|     | 0.00264                  | 35.6142    | 0.433512| 4.6510   | 0.0785807  | 29.4965  | 5.60573     |
| (b) | 0.024360                 | 39.130     | 4.59742 | 12.2093  | 1.57402    | 24.9431  | 8.01568     |
|     | 0.024356                 | 32.6816    | 3.32818 | 4.24807  | 0.54239    | 24.5482  | 4.25972     |
| (c) | 0.034775                 | 26.2807    | 3.7627  | 8.5555   | 1.1525     | 16.612   | 4.7536      |
|     | 0.076878                 | 28.4572    | 7.24541 | 3.6430   | 0.91001    | 17.831   | 2.4707      |

Table 2 Energy integrands: Production term, viscous dissipation and energy transfer due to Coriolis force are shown for $k_x = 1.2$ and $k_z = 5.626$ instabilities when $Re = 1228$ and $Ro = 0.15$.

## 5. Assessment with experiments

We examine the experimental results of Alfredsson & Persson (1989) to trace signatures of the Coriolis force driven instabilities found in our analysis. We find some signs of the presence of $Type-I$ and $Type-II$ instabilities from the flow visualization images reported by Alfredsson & Persson (1989) by comparing them to the predictions of linear stability theory. For instance, we observe that both $Type-I$ and $Type-II$ modes are unstable (see fig. 9 and 10(a)) at $Re = 590, Ro = 0.015$. Moreover, the most amplified spanwise wavenumber is $k_z = 4.6$ for the $Type-I$ (see fig. 10(b)) instability while that observed in the flow visualization image yields $k_z = 3.9$ which are reasonably close. If we set the spanwise and streamwise wavenumber equal to that observed in the flow visualization image (indicated by the cross in fig. 9.), we find that the associated mode shapes resemble the flow visualization image of Alfredsson & Persson (1989) in figure 11 (a)-(c). However, it is not possible to decipher from the flow visualization image as to which mode has a greater contribution to the instability observed in the experiment because the $x-z$ plane views at the mid plane of the instabilities look identical. Alternatively, it is possible that both instabilities co-exist and lead to "large-scale wavy fluctuations and twisting" of the roll-cells, observed in the experiments, at a downstream location due to non-linear effects. In order to distinguish, the dominant contribution from which mode, one must examine the roll-cells in the $x-y$ plane as shown in fig. 12 (a)-(b).The inclination to the horizontal of both, $Type-I$ and $Type-II$ modes is indicated which reveals that the latter has a stronger streamwise component. The precise contribution will however be determined by the make-up of the upstream disturbances. Video recordings of Alfredsson & Persson (1989) suggested the presence of waviness in the flow-field which propagates downstream at nearly half the undisturbed centreline velocity. The $\omega_r$ for $Type-I$ instability is 0.15134808 and for $Type-II$ instability is 0.16579894 and has a stronger streamwise velocity at the mid-plane compared to $Type-I$ instability. Indeed, the phase speed of $Type-I$ and $Type-II$ instabilities seem to match this description; an observation which might have escaped the Alfredsson & Persson (1989) because they considered only steady disturbances. We are therefore prompted to believe that $Type-I$ and $Type-II$ instabilities might both have co-existed in the experiments of Alfredsson & Persson (1989)(Alfredsson & Persson 1989).

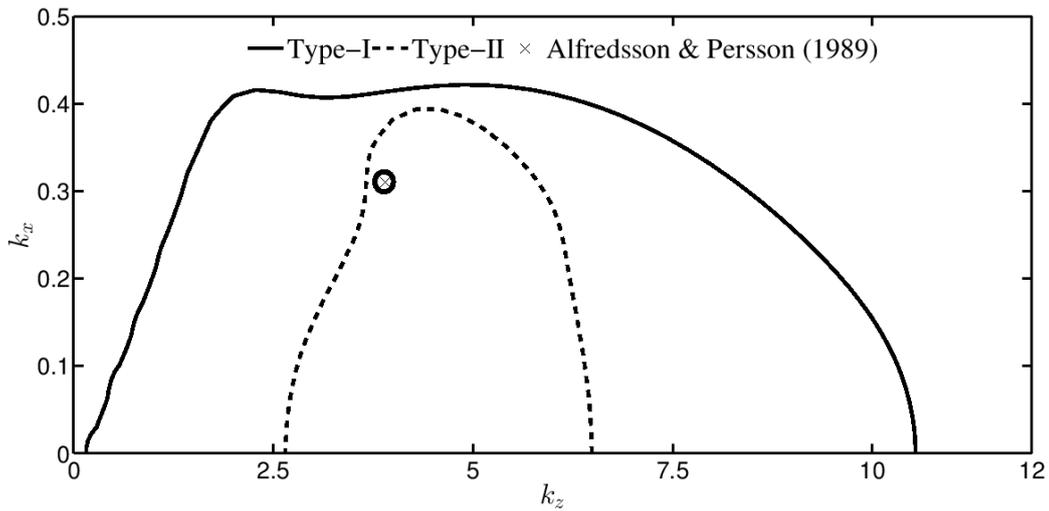

Figure 9. Comparison with experiment for $Re = 590.0$ and $Ro = 0.015$. Most unstable wave number of the disturbance: from experiment $k_z = 3.9$ and from linear stability theory $k_z = 4.6$.

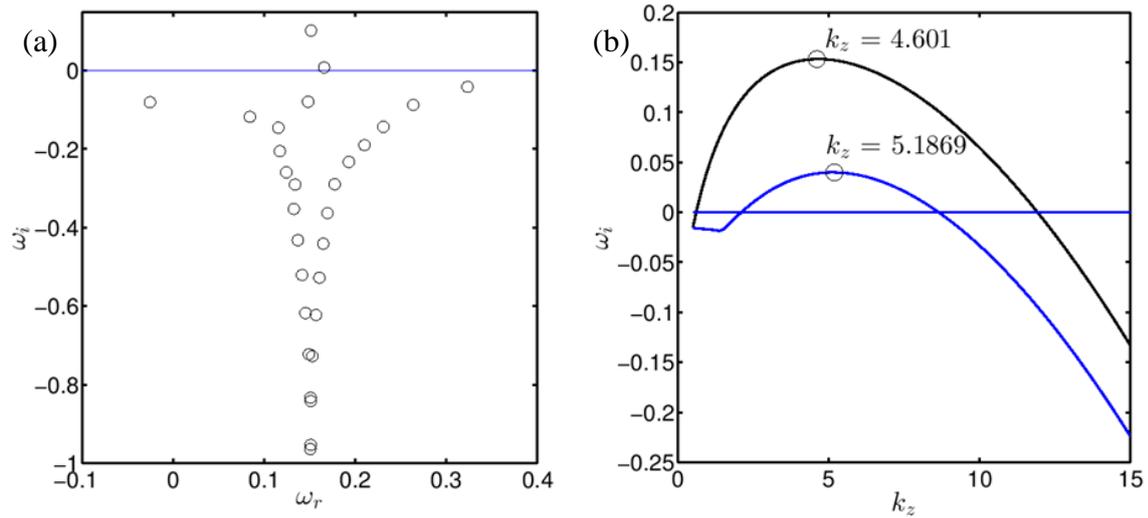

Figure 10. (a) The plot of eigen value spectrum with parameters for $Re = 590$ and $Ro = 0.015$ and $k_x = 0.15$, $k_z = 3.9$. The $\omega_r$ for $Type-I$ instability 0.15134808 is and for $Type-II$ instability is 0.16579894. (b) a plot of growth rate $(\omega_i - k_z)$ for the above parameters $(Re = 590.0, Ro = 0.015, k_x = 0.15)$.

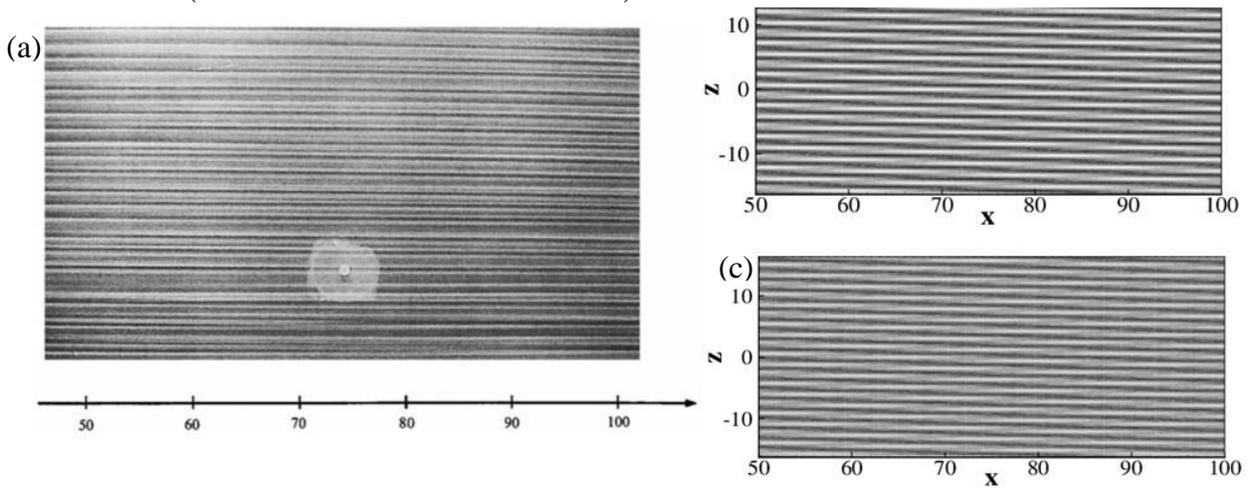

Figure 11. (a) Flow field in axial direction by Alfredsson & Persson (1989); (b) Flow field for $Type-I$ instability, and (c) Flow field for $Type-II$ instability. Reynolds number and rotation number for the flow is $Re = 590.0$ and $Ro = 0.015$.

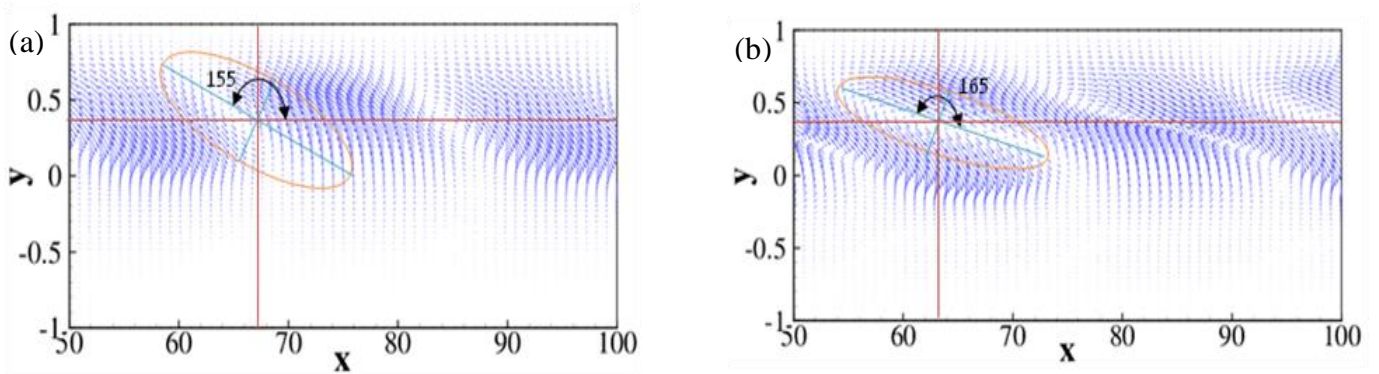

Figure 12. Structure of the roll cell formed due to (a) $Type-I$ instability & (b) $Type-II$ instability.

## 6. Microchannel design guideline

The regime diagram furnished below can help in designing a centrifugal microfluidic device, which need a strong instability to enhance the mixing process. The occurrence of four different types of instabilities mechanism in $Ro-Re$ plane is investigated for a wide range of spanwise and streamwise wavenumber. Figure 13(a) is drawn in $Ro-Re$ plane in order to assess clearly the influence of Reynolds and rotation number on the unstable modes.

$Type-I$ instability (which is well known in the literature) exists in very low Reynolds number, as low as $Re \sim 100$, which is confirmed with our regime diagram. However, $Type-II$ instability comes into existence at Reynolds number approximately $200$. $Type-III$ and $Type-IV$ instability, however, comes into existence at much higher Reynolds number. From the figure, we observe that four types of instability mechanism can coexist at higher Reynolds number and very low rotation number (as low as $Ro \sim 0.05$). We computed the penetration depth $(\delta)$ for different modes of instability in figure 13(b). The penetration depth for four different modes of the instability is defined as $\delta = \frac{2}{D_h} \int_{-1}^{p} |v| dy$, where $p$ is the location of the interface. This effectively calculates the modes responsible for efficient mixing in microscale for the centrifugal microfluidic device for the Reynolds number and rotation number considered. The typical time scale for mixing microchannel is several minutes and hence by exploiting the four modes of instability the efficient mixing time can be reduced considerably (Kuo & Jiang 2014). From the figure 13(b), we can see the penetration depth for normal velocity is highest for $Type-IV$ instability and lowest for $Type-I$ instability for four different penetration depth $p = 0.0, 0.5, 0.75, 1.0$. Thus, for typical microchannel flow, even if the growth rate of $Type-I$ instability is higher, in the short time scale, $Type-IV$ instability can be effective for micromixing process.

It is essential to mention that the slow-moving modes ($Type-III, IV$) of instability may play an important part in keeping the faster modes ($Type-I, II$) of instability at higher excited state and so, more unstable at higher spanwise wave number. The role played by $Type-III$ and $Type-IV$ instability for the considered flow may be to is to destabilize the flow configuration for much higher and wider range of $k_z$. First two modes are achieving stronger growth ($\omega_i$) due to the co-existence of unstable $Type-III, IV$ modes (see figure 14), which will effectively enhance the mixing effect at much smaller time scale. However, the phase speed and growth rate cannot be increased indefinitely. The most excited mode attains an optimal growth depending on $k_z$ and other parameters, when all four unstable modes are present.

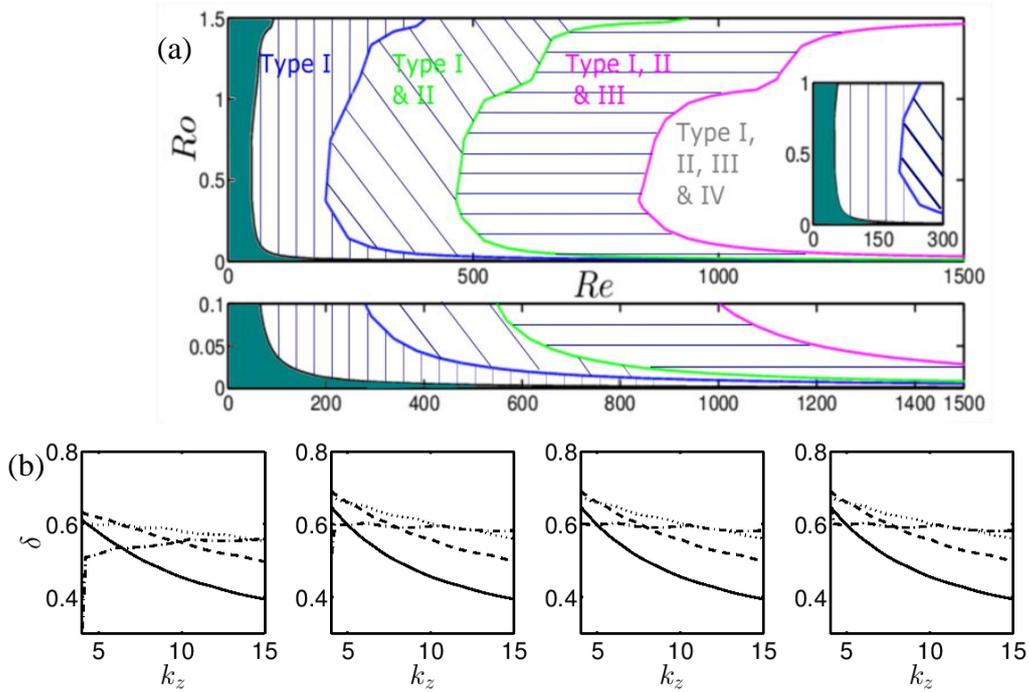

Figure 13. (a) Regime diagram, showing the presence of different modal instabilities in the $Re-Ro$ plane, and (b) penetration depth $(\delta)$ for different modes of instability. $Type-(I-IV)$ instability is represented by solid, dashed, dotted and dash dotted line respectively.

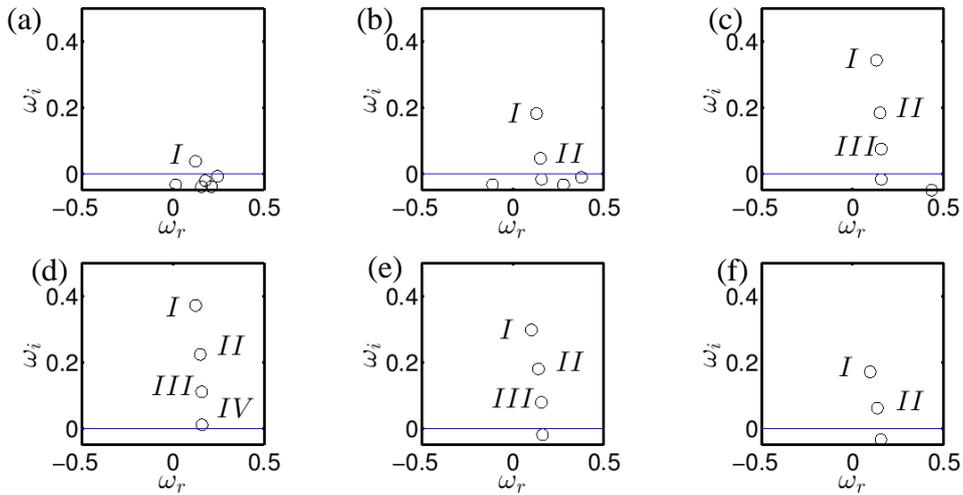

Figure 14. Occurrence of unstable modes for $k_z = 0.50, 1.5, 4.5, 6.5, 15.0$ and $20.0$ (figure (a) to (f)). Other parameters are $k_x = 0.15, Re = 1228.0, Ro = 0.15$ respectively.

# 7. Summary and conclusions

We have considered the steady fully developed plane Poiseuille flow through a channel subjected to system rotation about the spanwise direction. A temporal stability analysis has been performed using Chebyshev spectral collocation method to investigate the effect of spanwise disturbance on the flow thorough microfluidic channel. In particular, the influence of Coriolis force on the linear stability of the rotating microchannel flow susceptible to spanwise disturbances and the effect on micro mixing has been studied. The important observations from the present study are:

1. The results related to temporal stability analysis reveal the existence of four different types of instable modes, namely $Type\ I-IV$, at low streamwise wavenumber, originating from an energy transfer mechanism that hinges upon the Coriolis force. The regime diagram in $Ro-Re$ plane (figure 13(a)) supports the view that the four instabilities can coexist at high Reynolds numbers and small rotation numbers. Among the said modes, the two fastest growing modes, namely $Type\ I$ and $Type\ II$ mode, show prominent mode competition for certain range of parameters (figures 6(b)). To understand the effect of Coriolis force in the formation of roll-cells and the mode competition, a kinetic energy budget analysis was carried out. Results suggest that the rate of change in kinetic energy, that itself depends primarily on the relative contribution of the Coriolis force term to the kinetic energy of the roll-cells, is responsible for the observed competition of modes.

2. Investigation related to the structure of roll-cells that forms due to four different unstable eigenmodes (see figure 4) disclose that the $Type\ I$ instability shows regularly spaced roll cell structure. The structure of $Type\ II$ instability has a higher wavenumber is the wall normal direction and further flow-visualization studies might help identify the presence of this mode. However, the structures of $Type\ III$ and $Type\ IV$ instabilities are very resembling. Their structure displays twisted and wavy pattern in the mean flow direction of the roll cells.

3. We have compared our linear stability results with the experiment of Alfredsson & Persson (1989). Our results show very good agreement for most unstable spanwise wave number with the experimental results. A qualitative assessment of flow field in the axial direction has been performed. The considered flow simulated using eigen functions of $Type\ I$ and $Type\ II$ instabilities, shows identical character with that of the flow field observed in the experiment conducted by Alfredsson & Persson (1989). The results related to the marginal stability curve in laminar Plane Poiseuille flow with a spanwise system of rotation suggest a superposition of two different unstable eigenmodes.

Inferences drawn from the above observations would be far reaching. For instance, if an experiment is conducted in the above said system with a specified value of streamwise and spanwise disturbance, we can trigger *Type II* instability instead of *Type I* instability for which micro mixing may take place better. We also throw light on more realistic estimates of the *Type III* and *Type IV* instability and their contribution towards overall instability and mixing behaviour, an effect of which was not considered in the earlier investigations.